# A high resolution synchrotron x-ray powder diffraction study of the incommensurate modulation in the martensite phase of Ni$_2$MnGa: Evidence for nearly 7M modulation and phason broadening

Sanjay Singh[1,*], V. Petricek[2], Parasmani Rajput[3], Adrian H. Hill[3],

E. Suard[4], S. R. Barman[1], and Dhananjai Pandey[5]

[1] UGC-DAE Consortium for Scientific Research, Khandwa Road, Indore, 452001, India.

[2] Institute of Physics ASCR v.v.i., Department of Structure Analysis, Cukrovarnicka 10, 16200 Praha, Czech Republic

[3] European Synchrotron Radiation Facility, 6 rue Jules Horowitz, F-38000 Grenoble, France.

[4] Institut Laue-Langevin, BP 156, 38042 Grenoble Cedex 9, France

[5] School of Materials Science and Technology, Institute of Technology, Banaras Hindu University, Varanasi-221005, India.

## Abstract

The modulated structure of the martensite phase of Ni$_2$MnGa is revisited using high resolution synchrotron x-ray powder diffraction (SXRPD) measurements, which reveals higher order satellite reflections up to the 3$^{rd}$ order and phason broadening of the satellite peaks. The structure refinement, using the (3+1) dimensional superspace group approach, shows that the modulated structure of Ni$_2$MnGa can be described by orthorhombic superspace group

*Immm(00γ)s00* with lattice parameters a= 4.21861(2) Å, b= 5.54696(3) Å, and c= 4.18763(2) Å and an incommensurate modulation wave vector $\mathbf{q} = 0.43160(3)\mathbf{c}^* = (3/7+\delta)\mathbf{c}^*$, where $\delta = 0.00303(3)$ is the degree of incommensuration of the modulated structure. Additional satellite peak broadening, which could not be accounted for in terms of the anisotropic strain broadening based on a lattice parameter distribution, has been modeled in terms of phasons using fourth rank covariant strain tensor representation for incommensurate structures. The simulation of single crystal diffraction patterns from the refined structural parameters unambiguously reveals a rational approximant structure with 7M modulation. The inhomogeneous displacement of different atomic sites on account of incommensurate modulation and the presence of phason broadening clearly rule out the adaptive phase model proposed recently by *Kaufmann et al.*[1] and suggests that the modulation in $Ni_2MnGa$ originates from soft-mode phonons.

INTRODUCTION

Recent years have witnessed enormous interest in ferromagnetic shape memory alloys (FSMA) exhibiting extremely large magnetic field induced strain (MFIS) that is nearly an order of magnitude larger than the maximum strain generated in piezoelectric materials currently used by the industry for making miniaturized actuators for a host of applications [2]. FSMAs are therefore being visualised as better candidates for developing miniaturized magnetic actuators, but the extreme brittleness of some of the most promising FSMA compositions is a matter of concern[3, 4, 5, 6, 7]. FSMAs are magnetoelastic multiferroic materials exhibiting ferromagnetic and ferroelastic (martensitic) phase transitions with characteristic transition temperatures $T_C$ and $T_M$, respectively, with a very strong coupling between the magnetic and ferroelastic order parameters[8]. The martensitic transitions in FSMAs are displacive transitions resulting from a lattice variant deformation, usually called as Bain distortion, of the high temperature austenite phase leading to a low temperature martensite phase such that the large transformation strain is accommodated at the austenite-martensite interface (habit plane) by the formation of symmetry permitted martensite variants or ferroelastic domains (as distinct from the magnetic domains) through a lattice invariant deformation (achieved through twinning or faulting ) that leaves the habit plane undistorted and unrotated in an average sense at the microscopic scale [9]. Huge MFIS has been reported in the ferromagnetic martensite phase of the FMSAs with $T_c > T_M$ due to strong magnetoelastic coupling leading to magnetic field induced alignment of the ferroelastic domains (martensite variants) , and hence the magnetic moments, on account of low ferroelastic twinning energy as compared to the energy of magnetisation rotation[10, 11].

One of the most prominent ferromagnetic shape memory alloy systems is the Ni-Mn-Ga alloy, especially the nearly stoichiometric $Ni_2MnGa$ composition that shows about 10% magentic field induced strain in its low temperature martensite phase.[ 10, 11, 12] More recently, this alloy system has been shown to exhibit large magnetocaloric effect as well. [13] The room temperature structure (austenite phase) of $Ni_2MnGa$ is $L2_1$ type cubic in the Fm-3m space group that undergoes a ferromagnetic phase transition at $T_C \sim 370$ K, premartensitic transition to an incommensurate phase at $T_{PM}= 260$ K and a martensitic (ferroelastic) transition below $T_M= 210$K to a modulated structure.[14, 15] The structure of the low temperature martensite phase of $Ni_2MnGa$ has been extensively investigated using different diffraction techniques (x-ray, neutron and electron diffraction) for single crystal [16, 17, 18, 19, 20, 21], powder [19, 22, 23, 24] and epitaxial thin film [1] samples. However, the actual crystal structure and the nature of the modulation in the martensite phase, for which both commensurate and incommensurate modulations have been reported [16, 22, 23,24], are still controversial.

Webster et al. [15] studied the martensite structure of $Ni_2MnGa$ using neutron powder diffraction measurements and reported a tetragonal structure with c/a ratio of about 0.94. Martynov et al. reported five-layer modulated martensite structure (5M) on the basis of single crystal x-ray diffraction (XRD) data, where four satellites between the main austenite reflections were reported.[17] The first Rietveld refinement of the modulated structure of the martensite phase was carried out using medium resolution neutron powder diffraction data by Brown et al. [22] who concluded that the structure is orthorhombic in the Pnnm space group with 7M commensurate modulation. A subsequent Rietveld study using medium resolution x-ray powder diffraction data from a rotating anode based x-ray source also supported commensurate 7M modulation and Pnnm space group. [24] However, several workers have reported that the modulated phase of $Ni_2MnGa$ possesses an incommensurate structure. On the basis of Rietveld refinement in the (3+1) dimensional superspace using medium resolution laboratory source x-ray powder diffraction data, Righi et al. [23] reported incommensurate modulation with a modulation vector $\mathbf{q} = 0.4248(3)\mathbf{c}^*$ and concluded that the nearest rational approximant to the incommensurate phase is of 5M type, ie $\mathbf{q} = (2/5+\delta)\mathbf{c}^*$, where the incommensuration parameter δ is 0.0248(3), supposedly in agreement with electron diffraction studies by Pons et al. [25] and Chernenko et al. [26], in which four spots between the fundamental austenite reflections along 20$l$ reciprocal lattice rows were reported. However, subsequent single crystal diffraction studies [20] have revealed that the incommensurate phase exhibits six spots between the fundamental reflections suggesting a 7M type rational approximant for the stochiometric $Ni_2MnGa$ composition, which is in disagreement with Righi et al.'s conclusions. Furthermore, in the simulation of the single crystal diffraction patterns, Righi et al. considered second order satellite peaks, which were not directly observed in their medium resolution laboratory source XRD data and could only be generated from their structural model. Also their incommensuration parameter $\delta$ is relatively large. All these raise doubts about the correctness of their interpretation of the modulated structure being approximately 5M type.

It is evident from the foregoing that there is a twofold controversy about the nature of modulation in the martensite phase of $Ni_2MnGa$: (1) whether the modulation is of commensurate or incommensurate type and (2) whether the modulation is nearly (for incommensurate model) or exactly (for commensurate model) of 5M type or 7M type. Even single crystal diffraction measurements are not in agreement since evidence for both 5M and 7M like modulations with four and six superlattice spots between the main austenite peaks have been reported [17,20]. Apparently the composition of the sample has a very crucial role in deciding the nature of the modulation. This, therefore, necessitates to revisit the structure of martensite phase of stochiometric $Ni_2MnGa$ in detail. The powder diffraction patterns reported in literature have been recorded using moderate resolution powder diffractometers and laboratory X-ray sources or neutron sources. No attempt has been made to refine the modulated structure of the martensite phase of stochiometric $Ni_2MnGa$ using high resolution synchrotron x-ray powder diffraction data. The large MFIS in $Ni_2MnGa$ has been related to the modulated structure of the ferromagnetic martensite phase, which leads to lowering of the twinning stresses [10]. To understand the genesis of the large MFIS in $Ni_2MnGa$, it is therefore imperative to understand the structure and origin of the modulated phase.

We present here the results of Rietveld analysis of medium resolution neutron diffraction and very high resolution synchrotron x-ray powder diffraction (SXRPD) data using (3+1) dimensional superspace approach. The higher resolution and high intensity SXRPD data used in the present study shows not only well defined main reflections but also satellites reflection up to the third order, which enabled us to quantify the modulation wave vector precisely, as compared to the earlier low resolution XRD study using rotating anode data [23]. Furthermore, high resolution SXRPD data has enabled us to capture the signatures of additional broadening of the satellite peaks due to phasons, which could not be accounted for using Stephens model [27] of anisotropic peak broadening in commensurate structures and requires consideration of a 4$^{th}$ rank covariant tensor for incommensurate structures [28]. We also compare the simulated single crystal diffraction patterns of the incommensurate phase using first, second and third order satellites to confirm unambiguously that the structure of $Ni_2MnGa$ is 7M like, although incommensurate. The present results also indicate that the incommensurate modulation in stoichiometric $Ni_2MnGa$ cannot originate from the adaptive phase model [1] in view of the (1) significant mismatch between the calculated and observed peak positions of the superlattice reflections using commensurate 7M modulation, (2) dissimilar amplitudes of the atomic modulation functions for different atoms and (3) presence of large anisotropic peak broadening due to phasons.

METHODS

Polycrystalline $Ni_2MnGa$ was prepared by standard arc melting technique. The composition of the sample was confirmed to be stoichiometric $Ni_2MnGa$ using EDX analysis which gave a composition $Ni_{49.72}Mn_{25.24}Ga_{25.05}$ (~$Ni_{1.99}Mn_{1.01}Ga_{1.00}$, e/a~7.49). The premartensite transition temperature $T_{PM}=$ 261 K and martensitic transition temperature $T_M=$ 220K obtained from resistivity measurement [29] are consistent with $T_{PM}$ and $T_M$ reported by other workers for stoichiometric $Ni_2MnGa$ [16] The initial characterization results are given in Ref.[14].

The neutron diffraction measurements were performed at D2B beamline (ILL, Grenoble). -A vanadium cylinder was used as sample holder. The data were collected at 5K in the 2θ range of 10-160$^0$ in steps of 0.05$^0$ using a neutron wavelength of 1.59 Å in the high intensity mode. The minimum value of the full width at half maximum (FWHM) is around 0.07$^0$.

For the synchrotron XRD measurements, the same powder sample was sealed in a borosilicate capillary of 0.3mm diameter and the data was recorded at 90K at a wavelength of λ= 0.39993 Å in the 2θ range of 5-58$^0$ on the high- resolution powder diffractometer ID31 at ESRF, Grenoble. The resolution is given by the instrumental contribution to FWHM that is around 0.003$^0$ in 2θ. Le Bail and Rietveld analysis were performed using Fullprof [30] and JANA2006 software packages.[31]

RESULTS AND DISCUSSION

**Refinement using Commensurate Modulation:**

Rietveld refinement of commensurate modulated structure of the martensite phase of $Ni_2MnGa$ using medium resolution data has been carried out in the past.[22,24] In this section we proceed to show that medium resolution neutron powder diffraction data cannot capture the nature of incommensurate modulation in $Ni_2MnGa$. All the Bragg reflections in the neutron powder diffraction pattern of $Ni_2MnGa$ at 5K were well accounted for using orthorhombic space group *Pnnm* and the lattice parameters a≈ ( 1/√2) $a_{cubic}$, b≈ (7/√2) $a_{cubic}$ and c≈ $a_{cubic}$ where $a_{cubic}$ is the cell parameter of the cubic austenite phase [22]. The Rietveld fit is shown in Fig. 1. One peak at 2θ =40.5$^0$, which could not be account for, is due to aluminum in the cryofurnace wall (indicated by arrow in Fig. 1). The refined lattice parameters ($a$= 4.21796(6) Å, $b$= 29.2972(4) Å, $c$= 5.53492(6) Å) and the magnetic moment (3.13(7)$\mu_B$/f.u.) are in good agreement with the earlier neutron diffraction results.[22] From the values of lattice parameters it is evident that b≈(7/√2) $a_{cubic}$, where $a_{cubic}$ = 5.535 which indicates that the structure is 7-fold modulated (7M) in the <110> cubic direction in agreement with the observation of previous workers. [22, 24] While the neutron diffraction data analysis indicates that the structure of the martensite phase of $Ni_2MnGa$ is 7M modulated, the nature of the modulation, and in particular, whether it is commensurate or incommensurate, could not be settled unambiguously due to limited resolution of the data.

**Nature of Modulation and Evidence for Phason Broadening**

In order to determine the nature (commensurate vs incommensurate) of modulation, we use high resolution synchrotron x-ray diffraction data and analyse the data employing the (3+1)-D superspace approach.[32-34] In this approach the diffraction pattern is divided into two parts: (i) the main reflections corresponding to the basic structure and (ii) the satellite reflections arising out of the modulation and having weaker intensity as compared to the main reflections. All the main reflections due to the basic structure (Bain distorted structure) of $Ni_2MnGa$ could be indexed using an orthorhombic space group *Immm* with unit cell parameters a= 4.21853(2) Å, b=5.54667(2) Å and c=4.18754(1) Å which are similar to the earlier work [23]. After obtaining the unit cell parameters for the basic structure, the modulated structure was refined using the superspace group *Immm(00γ)s00* by (1) LeBail and (2) Rietveld techniques. This superspace group gives in $\mathbf{q} = 3/7 \mathbf{c}^*$ for the commensurate approximation in the supercell symmetry *Pnmn* for $t_0 = 0$ which is in accordance with [22].

We first present the results of Le Bail refinement for the commensurate and incommensuarte models in which only cell, profile (pseudo-Voigt), zero shift and background parameters were refined. Initially commensurate value of the modulation vector [$\mathbf{q} = 3/7\mathbf{c}^*$] and only first order satellites (hkl±1) were considered in the LeBail refinement. But, this model was unable to fit the satellite reflections as the calculated peak positions were shifted away from the observed ones (see Fig. 2 (a)). This shows the inadequacy of low resolution powder diffraction data in capturing the signatures of the failure of the commensurate modulation model of the structure like that presented in the previous section. We also tried the incommensurate value of the modulation vector reported in Ref.[23] (*q= 0.4248)*, but it led to even worse fit with higher GOF=4.82 compared to that (GOF= 4.04) for the commensurate modulation with $\mathbf{q} = 3/7\mathbf{c}^*$. This indicates that the modulation vector, as reported in [6], cannot account for the satellite peak positions. Finally we refined the modulation vector also and this led to significantly better fit between the observed and the calculated peak positions (see Fig. 2(b)) with lower GOF (= 3.69) for an incommensurate value of the modulation vector $\mathbf{q}$= 0.43154(3) $\mathbf{c}^*$.    So far we considered only the first order satellite reflections (hkl ± 1) in the refinements following Righi et al. [23] and it was possible to index majority of the satellite peaks. However, several of the satellite reflections with rather low intensities could not be accounted for using first order reflections only as shown in Fig. 3(a). Consideration of second order satellites (hkl ±2) in the LeBail refinement could index most of these low intensity reflections very well, as can be seen from Fig. 3(b). This confirmed the presence of second order satellites in our high resolution SXRPD patterns which were not discernible in the laboratory XRPD patterns of Righi et al [23]. It is interesting to note that consideration of even the second order satellites could not index some very weak reflections as can be seen in Fig 3(b) at 2θ ~10.03 deg. Accordingly, we considered third order satellites in our refinements and this led to the identification of the small peak at 2θ ~10.03 deg as a third order satellite as shown in Fig 3(c) (indicated by red arrow). It is worth mentioning here that there is no previous report on the observation of second and third order satellites in the powder diffraction pattern of the martensite phase of $Ni_2MnGa$ and it shows the significance of the higher resolution data used in the present work.

However, after including the third order satellites, it was noticed that the anisotropic peak broadening function in terms of Stephens [27] formalism, as used for 3D periodic crystals, cannot fully account for peak broadening of satellite reflections. It is well known that the incommensurate modulated structures may exhibit additional broadening of the satellite peaks due to phasons [28]. A fourth-rank covariant strain tensor based formalism with a distribution of the strain-tensor components implemented in JANA2006 [31] was therefore considered in the LeBail refinements. Ten strain components allowed by the symmetry were all considered in the refinement. This led to significantly better fit between the observed and calculated peak profiles (Fig 4(b)) with a decrease in the GOF from a value of 3.09 to 2.60. The values of the refined components of the strain tensor are given in Table I. The higher magnitude of strain for st2011 and st0022 (these coefficients are connected with the terms $h^2lm$ and $l^2m^2$, respectively – for more details see [27] and [28]) confirms the presence of phasons, not considered in any of the previous refinements [23, 35]. It is interesting to note that inelastic neutron scattering studies on the martensite phase of $Ni_2MnGa$ have revealed well-defined phasons associated with CDW[36]. Our LeBail refinements using fourth rank covariant strain tensor provide the first experimental evidence of such phason broadening resulting from the fluctuations in the incommensurate modulation vector in high resolution SXRPD patterns.

Having established from LeBail refinements the incommensurate nature of the modulated structure of the martensite phase of $Ni_2MnGa$, Rietveld refinements were performed. In the Rietveld refinement of the modulated structure, the deviation $\mathbf{u}(\bar{x}_4)$ from the average atomic position $x_i$ (i = 1 to 3) is refined. The deviation is described in terms of the superposition of atomic modulation functions, which are assumed to be harmonic functions of internal coordinate $x_4$. The actual atomic positions ($x_i$) in the incommensurate phase are thus given as:

$$\mathbf{x}(\bar{x}_4) = \bar{\mathbf{x}} + \mathbf{u}(\bar{x}_4) \quad \text{(i)}$$

$$\mathbf{u}(\bar{x}_4) = \sum_{1}^{\infty} [A_n \sin(2\pi n \bar{x}_4) + B_n \cos(2\pi n \bar{x}_4)] \quad \text{(ii)}$$

where $\bar{x}_i$ $(i = 1,2,3)$ is the general atomic position in the basic structure, $A_n$ and $B_n$ are the amplitudes of the displacement modulation and $n$ is the order of the Fourier series.

The deviation $\mathbf{u}(\bar{x}_4)$ from the basic atomic positions $x_i$ (i = 1 to 3), isotropic ADP ($U_{iso}$) and all other parameters considered in the Rietveld technique were refined taking into account the third order satellites and phason broadening. The comparison of the observed and calculated profiles is shown in Fig.5. Both the second and third order satellites could be fitted extremely well using incommensurate modulation model, as can be seen from one of the insets to Fig.5(a). The other inset in Fig. 5 (a) and Fig. 5 (b) illustrate the quality of the Rietveld fits in the lower and higher 2θ ranges. The refined basic cell parameters a= 4.21861(2) Å, b= 5.54696(3) Å, and c= 4.18763(2) Å and modulation vector $\mathbf{q} = 0.43160(3)\mathbf{c}^*$ are close to the values obtained by LeBail refinement. The refined atomic positions and the amplitudes of the modulation functions of the incommensurately modulated phase are listed in Table II. The average first nearest neighbor inter atomic distances obtained in the present study are given in Table III. No unphysically short interatomic distances for Ni-Ga and Ni-Mn are observed, unlike those for the commensurate modulation [22]. The Mn-Ga average distance (2.77355(6) Å) obtained here is in excellent agreement with the value (2.780(6)) reported by a x-ray absorption fine structure (EXAFS) study [37]. Moreover, besides Mn-Ga, other distances listed in Table III are also close to the EXAFS values. This excellent agreement between the interatomic distances obtained by the XRD technique that probes the spatially averaged long range structure and EXAFS that probes the local structure shows that local distortion of the structure is rather negligible. However, the Ni-Mn and Ni-Ga distances are smaller than the values derived from the atomic radii (Table III). The shorter Ni-Ga and Ni-Mn distances may be attributed to Ni-Ga and Ni-Mn hybridization [37].

If one uses only first order harmonic waves, the modulations of the atomic positions are allowed only along the x direction for all the atoms due to the symmetry restrictions [23]. However, higher order modulation waves (second and third) also allow modulations in the y direction for Ni atom and in the z direction for all other atoms. The displacements along the z direction are found to be within the standard deviations as can be seen from Table II. Further, significant displacement along y direction exists only for the Ni atoms. But more significantly, the atomic displacements along the x direction are different for each atom (Table II). The displacement amplitude ($A_1$) for the Mn, Ga and Ni atoms are 0.0665(12), 0.0657(9) and 0.0618(9), respectively (see Table II). Thus the largest $A_1$ is observed for Mn and it is lowest for Ni. The values of $A_1$ reported here are significantly different from earlier XRD results based on consideration of the first order satellites only [23] where these displacement were 0.066(8), 0.070(6) and 0.072(6) for Mn, Ga and Ni, respectively, in the x-direction only.

**Origin of the Modulated Structure**

There are essentially two main theories for the formation of the modulated structure of the martensite phase in $Ni_2MnGa$: the adaptive phase model and the soft-phonon mode based displacive modulation model. In the adaptive phase model [1], the modulated structure is visualized as a nanotwinned state of the Bain distorted phase, which maintains the invariance of the habit plane between the austenite and the martensite phases. In the soft phonon model, the origin of modulation has been related to a TA2 soft phonon mode in the transverse acoustic branch along [110] direction of the austenite phase [38-41], which has been supported by the observation of change in the modulation period leading to a premartensite phase before the final structural transition to the martensite phase [39]. The instability of the TA2 phonon is related to a long-range anomalous contribution to the phonon frequency due to electronic screening [42]. It is possible to make a choice between the adaptive phase and soft mode models from a knowledge of the amplitudes of displacive modulations for the different atomic sites, since they are required to be identical for the former but may be dissimilar for the latter.[35]. The considerably dissimilar amplitudes of modulation for the different atomic sites together with atomic displacements in different directions (Table II) clearly indicate that the modulations in stoichiometric $Ni_2MnGa$ cannot be explained in terms of the adaptive phase model but may arise due to soft phonon modes.[35] The observation of charge density wave in $Ni_2MnGa$ also supports the present finding and indicates that the modulation is driven by soft- phonon modes.[21] Thus our present results raise doubts about the validity of adaptive phase model.[1]

The shift of some of the x-ray powder diffraction peaks with the respect to the ideal commensurate positions predicted by adaptive phase model has been attributed by Kaufmann et al.[1] to the presence of stacking faults. Stacking faults are known to broaden and shift the peaks as discussed in the context of non- magnetic shape memory alloys by Kabra et al.[43] However, no explanation exists as to why the shifts should be only for the satellite reflections, as the stacking faults are known to affect the main peaks also. More importantly, simulation of the diffraction patterns from the nanotwinned $Ni_2MnGa$ structure with stacking faults has revealed that the observed peak shifts cannot be attributed to stacking faults [16]. Thus, the adaptive phase model fails to explain the incommensurate modulated structure even if the presence of stacking faults is invoked.

It is worth mentioning here that although the intensity of the second and third order satellites is comparatively less, it has played an important role in obtaining the modulation amplitudes with greater accuracy and hence in the rejection of the adaptive phase model. Righi et al. [23] could not observe second and third order satellites as they used lower intensity and higher background laboratory source XRD data.

**The Periodicity of the Rational Approximant Structure**

We have also simulated the single crystal diffraction patterns along [010] zone axis to compare our results with those obtained by earlier workers [16, 20, 23]. The simulation was carried out using the JANA2006 package. Fig. 6(a) depicts the reciprocal space section calculated using $1^{st}$ order satellites only, as per the results of Righi et al. [23] It is evident that only two satellites can be observed between the main reflections in this case and therefore the diffraction pattern indicates a 3M like modulation. Although $2^{nd}$ order satellites were also considered by Righi et al in their simulations of the single crystal diffraction patterns, they had not observed $2^{nd}$ order satellites in their diffraction patterns. In the present study we have found unambiguous evidence for the second order satellite reflections, and their consideration in the simulation gives rise to 4 satellites between the main reflections in the reciprocal space (Fig. 6(b)). Since we have observed $3^{rd}$ order satellites also, we simulated the reciprocal space including these satellites and the results are shown in Fig. 6(c). There are evidently six satellites between the main reflections, which is in good agreement with the electron diffraction patterns of Fukuda et al. [20]. Our simulations thus reveal that the labels like 3M, 5M, 7M type modulated structure of the martensite phase are fraught with ambiguities, as it depends on the resolution of the diffraction technique used.. In fact, between the two main diffraction spots of the I centered cell (say, (hkl0) with h+k+l=2n and (hkl+2,0)), one could in principle observe not only six satellites but even 13 satellites including the main reflection (h,k,l+1,0) which is systematically extinct in the original cell. However, the intensities of satellites, having order higher than 3, are too weak to be discernible. Since satellites upto the $3^{rd}$ order are discernible in our high resolution SXRPD data, we propose that the 7M modulation is the best plausible description of the rational approximant of the incommensurate modulated phase in agreement with single crystal diffraction results by other workers [20, 39].

From the present (3+1)-D incommensurate model, we have derived the 3D rational approximant superstructure shown in Fig. 7 taking q≈3/7. It involves seven unit cells of the basic structure along the c-axis and can be described by the *Pnmn* space group in agreement with the findings of Brown et al. [22] and Ranjan et al.[24] The wave-like displacement of atoms is ascribed to the structural modulation. It is interesting to note from the figure that the displacement of all the atoms is in the same phase, as expected for a rational approximant. The atomic position, space group and lattice parameters for 3D model are given in Table IV. Our results clearly rule out the 5M description [23] for the rational approximant of the incommensurate phase of $Ni_2MnGa$.

**CONCLUSIONS**

In conclusions, we have presented results of (3+1) D superspace Le Bail and Rietveld refinements of the modulated structure of the martensinte phase of stoichiometric $Ni_2MnGa$ ferromagnetic shape memory alloy using high resolution synchrotron x-ray powder diffraction (SXRPD) patterns. Our results confirm the incommensurate nature of the modulation and presence of phason broadening of the satellite peaks. Observation of higher order satellites up to the 3$^{rd}$ has enabled us to simulate the single crystal x-ray diffraction patterns, which reveal six satellite between the main diffraction unambiguously. This conclusively rejects the 5M like rational approximant structure of the martensite phase and confirms the 7M like modulation. Presence of higher order satellites due to the higher resolution of the SXRPD data has enabled us to determine the modulation wave vector precisely as $\mathbf{q} = (3/7 + \delta)\mathbf{c}^*$ where the incommensuration parameter δ =0.00303(3) and capture the atomic displacements not only along the x direction but also along the y direction. The inhomogeneous nature of the atomic displacement rule out the adaptive phase model as a possible mechanism for the origin of the modulated structure and suggests the soft phonon mode as the most plausible mechanism of modulation.

ACKNOWLEDGMENTS

A. Fitch is thanked for useful discussion and support during the SXRPD data collection. Fundings from; Department of Science and Technology, Government of India through S. N. Bose National Centre for Basic Sciences; and Institut Laue-Langevin, France are gratefully acknowledged. S.S. thanks the Council for Scientific and Industrial Research for research fellowship. DP thanks the Science and Engineering Research Board of India for the award of J.C. Bose National Fellowship.

*Present address: Experimentalphysik, University Duisburg- Essen, D-47048 Duisburg, Germany.

**Figures:**

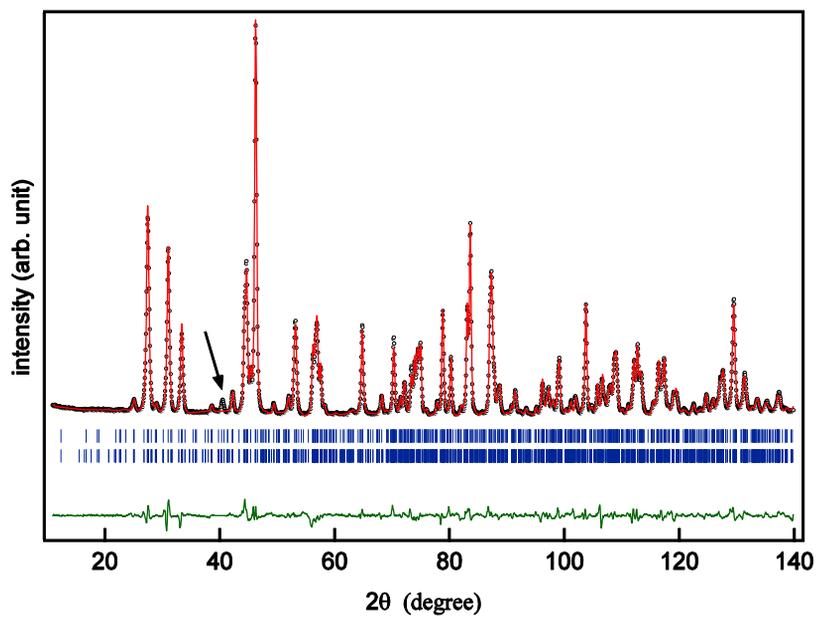

Fig. 1. (color online) Rietveld fitting for the neutron powder diffraction pattern of the martensite phase of $Ni_2MnGa$ at 5K. The experimental data, fitted curves and the residue are shown by dots (black), continuous line (red line) and the bottom most plot (green line), respectively. The upper and lower rows of ticks (blue) represent the nuclear and magnetic Bragg peak positions, respectively. The arrow indicates a peak due to Aluminum (see text).

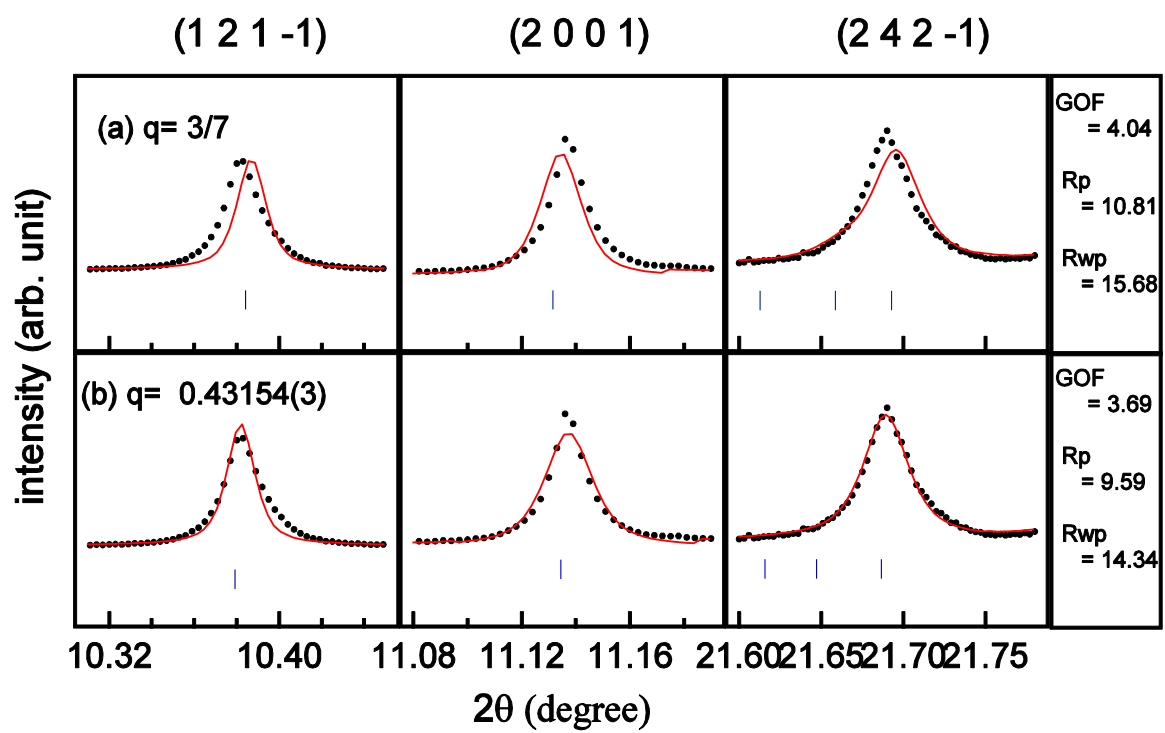

Fig. 2. (color online) The Le Bail fitting for some satellite reflections in the SXRPD pattern of the martensite phase (90K) of Ni$_2$MnGa for (a) **q**= (3/7) **c**\*(b) **q**= 0.43154(3)**c**\*. The dots (black) and continuous line (red) represent the observed and calculated profiles, while the vertical tick mark represents the Bragg position. The fitting parameters, GOF, Rp, Rwp, are also shown in the last column.

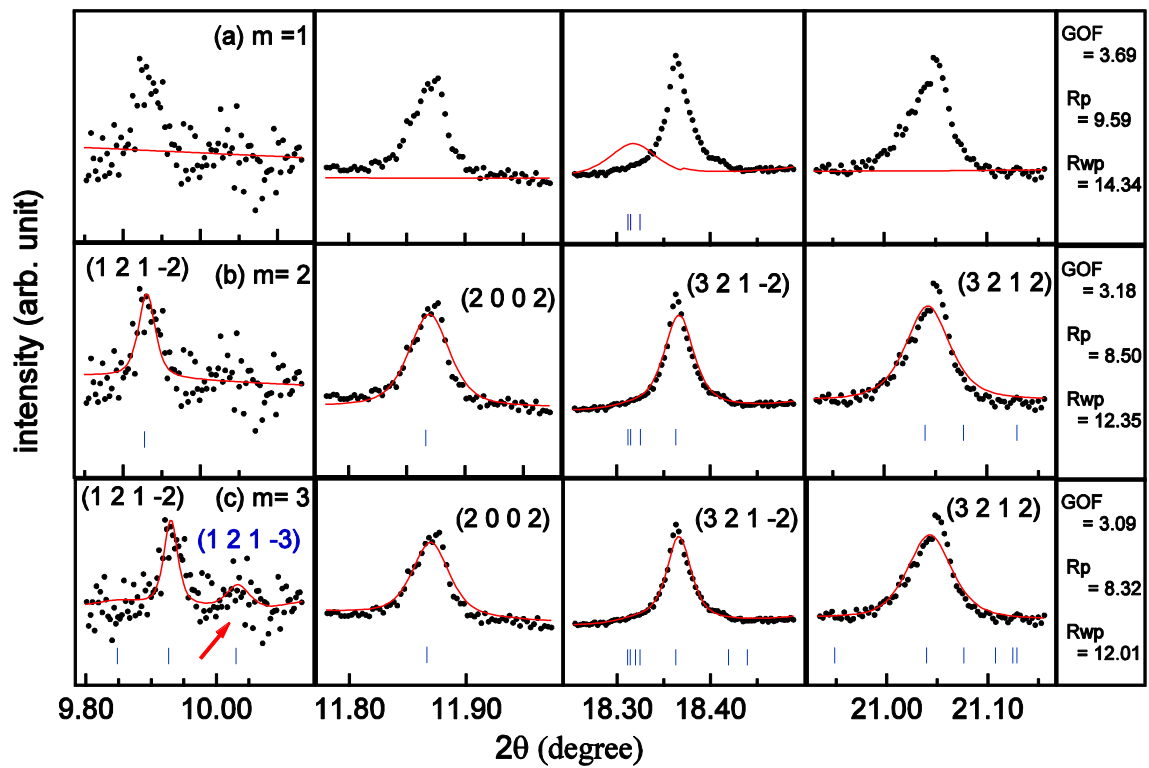

Fig. 3. (color online) The Le Bail fitting of some satellite peaks in the SXRPD pattern of the martensite phase (90K) of Ni$_2$MnGa considering (a) only first order satellites (hkl ±1) (b) first and second order satellites (hkl ± 2) and (c) first, second and third order satellites (hkl ± 3). The third order is indicated by red arrow. The symbols have the same meaning as in Fig. 2.

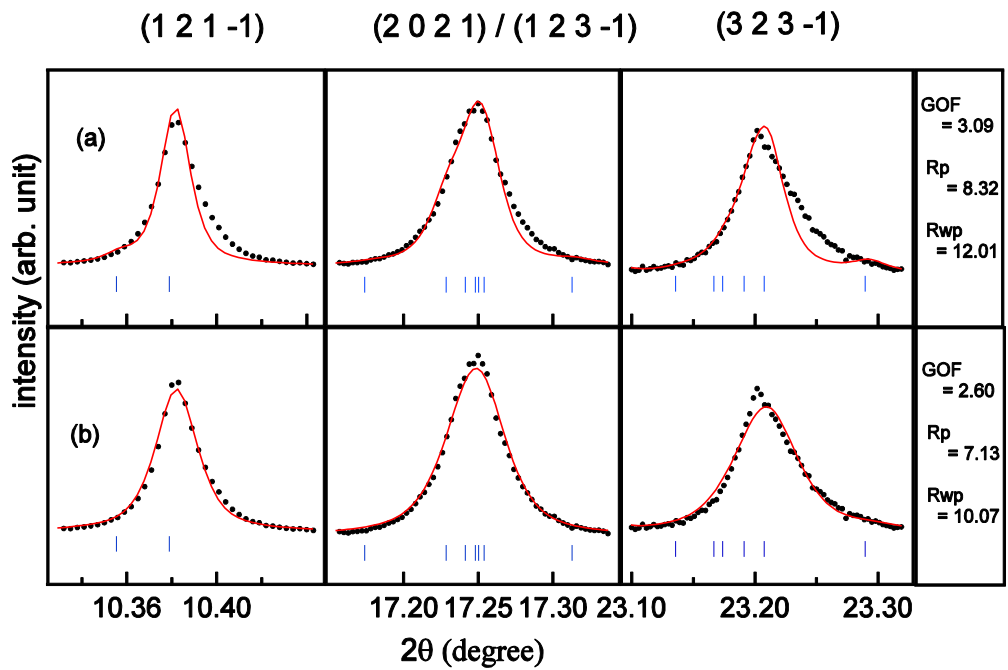

Fig.4. (color online) The Le Bail fitting of some satellite reflections (a) without considering and (b) after considering the fourth-rank covariant tensor to represent phason broadening. The symbols have the same meaning as in Fig. 2.

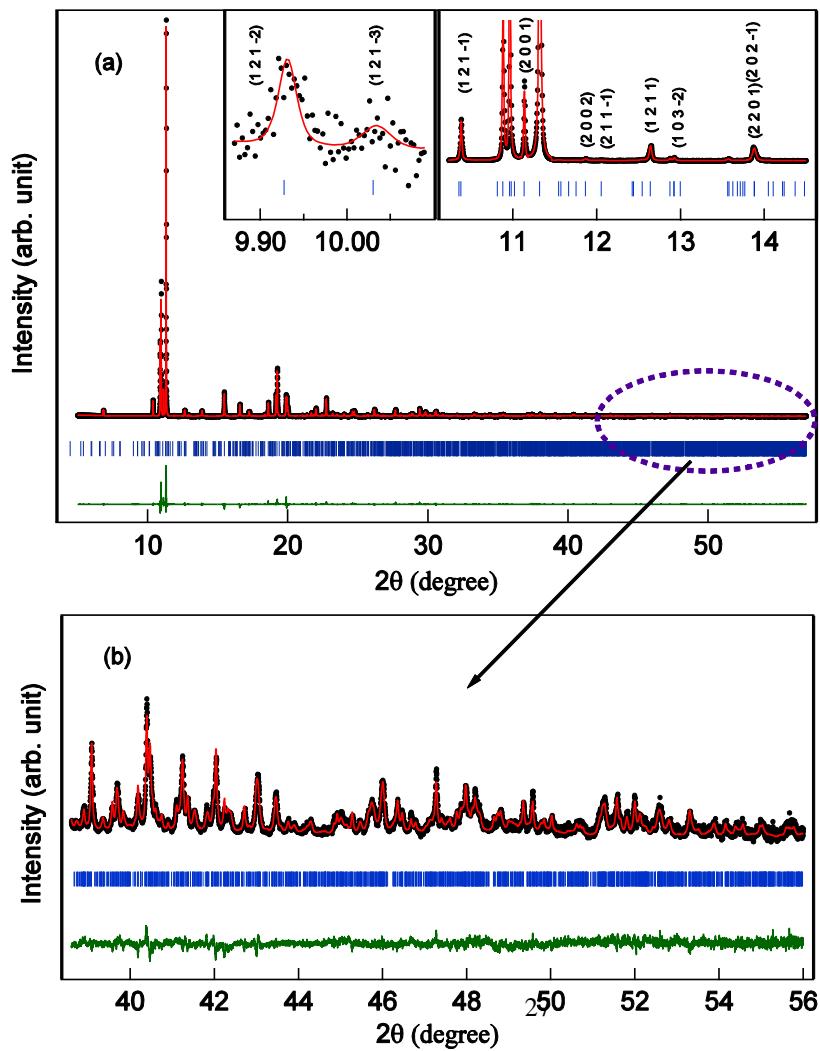

Fig. 5. (color online) (a) Rietveld fitting for the incommensurate modulated martensite phase of $Ni_2MnGa$ at 90K. The experimental data, fitted curve and the residue are shown by dots (black), continuous line (red) and bottom most plot (green), respectively. The tick marks (blue) represent the Bragg peak positions. The inset in (a) on the left shows the fit for $2^{nd}$ and $3^{rd}$ order satellites. while the second inset on the right side shows the fit for the main peak region in an expanded scale. (b) The fitting for the high angle reflections is shown on an expanded scale for higher $2\theta$ range ($39^0 - 56^0$).

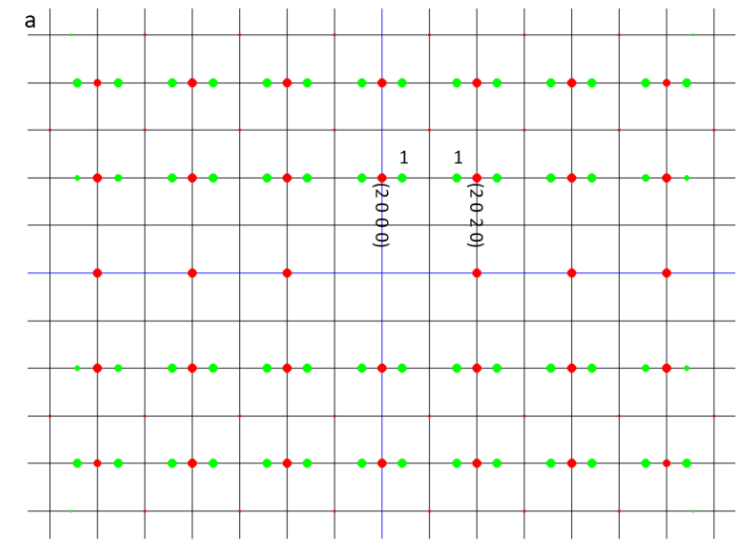

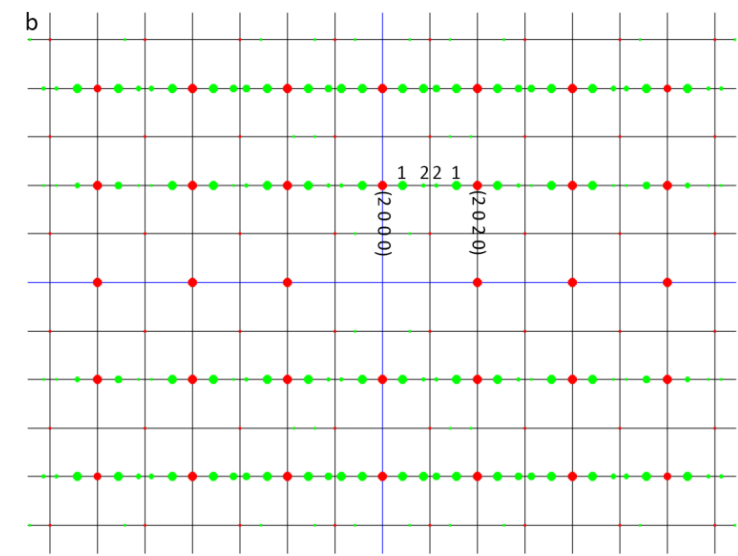

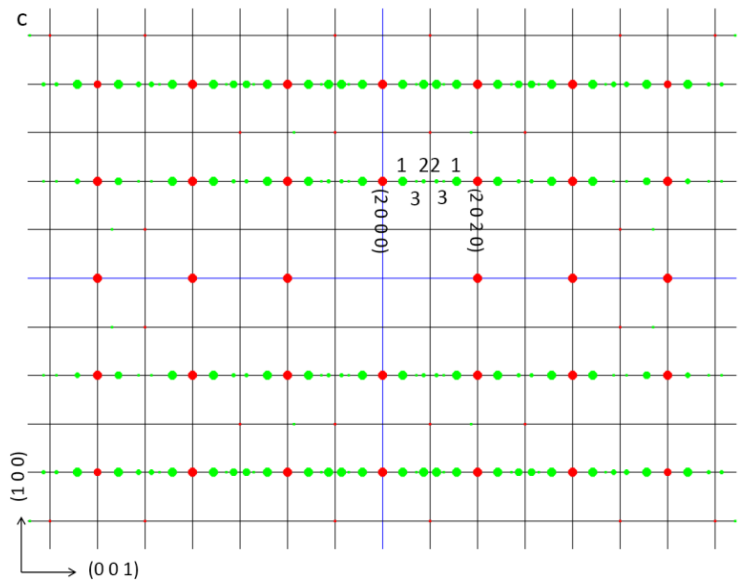

Fig. 6. Simulation of the (010) section of the reciprocal space of the incommensurate modulated structure using satellite peaks upto (a) $1^{st}$ order, (b) $2^{nd}$ order and (c) $3^{rd}$ order. Note the 3M, 5M and 7M like patterns for (a), (b) and (c), respectively. Red spots are the main reflections and Green spots are satellites. Numbers 1, 2 and 3 represent the order of the satellites.

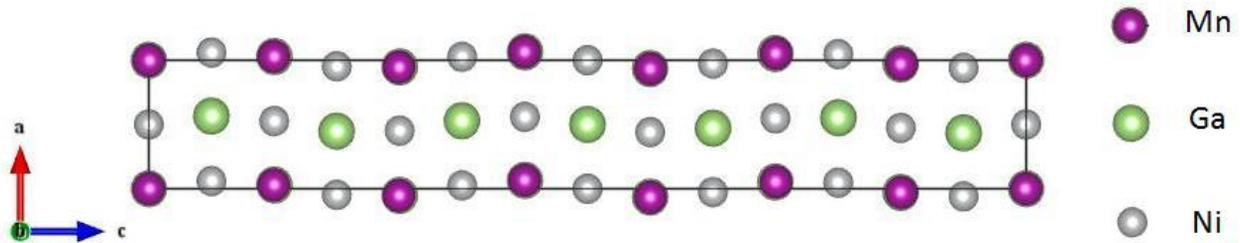

Fig.7. (Color online) the modulated orthorhombic unit cell of $Ni_2MnGa$ martensite phase projected in the a−c plane highlighting the atomic position modulation in its rational approximant 7M structure.

**Tables:**

TableI: The refined values of the strain components of the fourth-rank covariance tensor.

| Strain Component | Value | Strain Component | Value |
|---|---|---|---|
| st4000 | 3.7 | st2002 | 11.6 |
| st0211 | 10 | st2200 | 13.4 |
| st2020 | 40.6 | st2011 | 20.4 |
| st0400 | 28.2 | st0022 | 29.4 |

**TableII :**

Atomic positions $(x, y, z)$, atomic displacement parameter $(U_{iso})$ and amplitudes $(A_1, B_1, A_2, B_2, A_3, B_3)$ of the modulation function of the incommensurate modulated martensite phase of $Ni_2MnGa$.

| Atom | Wyckoff position | Modulation amplitude | x | y | z | $U_{iso}(Å^2)$ |
|---|---|---|---|---|---|---|
| Ga1 | 2d | | 0 | 0.5 | 0 | 0.0037(5) |
| | | $A_1$ | 0.0657(9) | 0 | 0 | |
| | | $B_1$ | 0 | 0 | 0 | |
| | | $A_2$ | 0 | 0 | 0.001(3) | |
| | | $B_2$ | 0 | 0 | 0 | |
| | | $A_3$ | -0.005(2) | 0 | 0 | |
| | | $B_3$ | 0 | 0 | 0 | |
| Mn1 | 2a | | 0 | 0 | 0 | 0.0030(6) |
| | | $A_1$ | 0.0665(12) | 0 | 0 | |
| | | $B_1$ | 0 | 0 | 0 | |
| | | $A_2$ | 0 | 0 | -0.001(4) | |
| | | $B_2$ | 0 | 0 | 0 | |
| | | $A_3$ | -0.001(3) | 0 | 0 | |
| | | $B_3$ | 0 | 0 | 0 | |
| Ni1 | 4h | | 0.5 | 0.25 | 0 | 0.0013(4) |
| | | $A_1$ | 0.0618(9) | 0 | 0 | |
| | | $B_1$ | 0 | 0 | 0 | |
| | | $A_2$ | 0 | 0 | 0.000(3) | |
| | | $B_2$ | 0 | -0.0029(7) | 0 | |
| | | $A_3$ | -0.004(2) | 0 | 0 | |
| | | $B_3$ | 0 | 0 | 0 | |

**Table III:**

The average interatomic bond distances for the 7M phase of $Ni_2MnGa$ and expected interatomic distances, assuming close packed hard sphere model, obtained from the atomic radii (Ni: 1.25Å, Mn: 1.37Å, Ga: 1.53Å).

| Atom1-Atom2 | Average (Å) | Minimum (Å) | Maximum (Å) | Distance calculated from atomic radii (Å) |
|:---:|:---:|:---:|:---:|:---:|
| Ni1-Ga1 | 2.523(8) | 2.509(12) | 2.535(5) | 2.78 |
| Ni1-Mn1 | 2.523(9) | 2.509(12) | 2.536(7) | 2.62 |
| Ni1-Ni1 | 2.773(3) | 2.741(6) | 2.806(6) | 2.5 |
| Mn1-Ga1 | 2.77352(7) | 2.77347(13) | 2.77356(13) | 2.9 |
| Mn1-Mn1 | 4.069(10) | 3.894(14) | 4.253(14) | 2.74 |
| Ga1-Ga1 | 4.069(7) | 3.919(7) | 4.239(11) | 3.06 |

**TableIV:**

Lattice parameters, atomic positions (x, y, z) and atomic displacement parameter (Uiso) for 7M commensurate modulated structure of the martensite phase of $Ni_2MnGa$.

| Crystal system | | Orthorhombic | | | |
|---|---|---|---|---|---|
| Space group | | *Pnmn* | | | |
| Cell (Å) | a= 4.21861(2), | | b= 5.54696(3), | | 29.31344(2) |
| Atom | Wyckoff position | $x$ | $y$ | $z$ | $U_{iso}$ (Å$^2$) |
| Ga1 | 2a | 0 | 0.5 | 0 | 0.0037(3) |
| Ga2 | 4g | 0.023(6) | 0.5 | 0.143(2) | 0.0037(3) |
| Ga3 | 4g | 0.951(7) | 0.5 | 0.286(4) | 0.0037(3) |
| Ga4 | 4g | 0.068(4) | 0.5 | 0.428(5) | 0.0037(3) |
| Mn1 | 2b | 0 | 0 | 0 | 0.0032(5) |
| Mn2 | 4g | 0.028(9) | 0 | 0.1429(14) | 0.0032(5) |
| Mn3 | 4g | 0.948(12) | 0 | 0.286(7) | 0.0032(5) |
| Mn4 | 4g | 0.065(6) | 0 | 0.429(10) | 0.0032(5) |
| Ni1 | 4f | 0.5 | 0.247(17) | 0 | 0.0013(2) |
| Ni2 | 8h | 0.523(3) | 0.248(15) | 0.1429(14) | 0.0013(2) |
| Ni3 | 8h | 0.454(4) | 0.251(11) | 0.286(2) | 0.0013(2) |
| Ni4 | 8h | 0.564(2) | 0.253(5) | 0.429(3) | 0.0013(2) |